\documentclass[prb,amsmath,twocolumn]{revtex4}
\usepackage{graphicx}
\usepackage{amssymb}
\usepackage[percent]{overpic}
\begin{document}
\title{Theory of triangular lattice quasi-one-dimensional charge-transfer solids}  
\author{R. T. Clay}
\email{r.t.clay@msstate.edu}
\affiliation{Department of Physics \& Astronomy and HPC$^2$ Center for Computational Sciences, Mississippi State University, Mississippi State, MS 39762}
\author{N. Gomes}
\author{S. Mazumdar}
\affiliation{Department of Physics, University of Arizona, Tucson, AZ 85721}
\date{today}
\vskip 1pc
\begin{abstract}
  Recent investigations of the magnetic properties and the discovery
  of superconductivity in quasi-one-dimensional triangular lattice
  organic charge-transfer solids have indicated the severe limitations
  of the effective $\frac{1}{2}$-filled band Hubbard model for these
  and related systems.  We present computational studies within the
  $\frac{1}{4}$-filled band Hubbard model for these highly anisotropic
  systems.  Individual organic monomer molecules, and not their
  dimers, constitute the sites of the Hamiltonian within our theory.
  We find enhancement of the long-range component of superconducting
  pairing correlations by the Hubbard repulsive interaction $U$ for
  band parameters corresponding to
  $\kappa$-(BEDT-TTF)$_2$CF$_3$SO$_3$, which is superconducting under
  moderate pressure.  We find significantly weaker superconducting
  pairing at realistic values of $U$ in
  $\kappa$-(BEDT-TTF)$_2$B(CN)$_4$, and we ascribe the experimentally
  observed transition to a spin-gapped insulator to the formation of a
  paired-electron crystal. We make the testable prediction that the
  spin gap will be accompanied by charge-ordering and period doubling
  in two lattice directions.  The weaker tendency to superconductivity
  in $\kappa$-(BEDT-TTF)$_2$B(CN)$_4$ compared to
  $\kappa$-(BEDT-TTF)$_2$CF$_3$SO$_3$ is consistent with the more
  one-dimensional character of the former.  Pressure-induced
  superconductivity is, however, conceivable. The overall results
  support a valence bond theory of superconductivity we have proposed
  recently.
\end{abstract}
\maketitle

\section{Introduction}

Determination of the mechanism of correlated-electron
superconductivity (SC) continues to be a formidable challenge, even
after three decades of intensive investigations following the
discovery of the phenomenon in doped copper oxides. Discoveries of
many other exotic superconductors subsequently have only added to its
mystique.  One opinion, shared by many scientists, is that the
mechanisms of correlated-electron SC in apparently different families
are related
\cite{McKenzie97a,Yanase03a,Dagotto05a,Steglich16a,Clay19a}.  Among
the many families of materials that have attracted attention in this
context are organic charge-transfer solids (CTS), SC in which was
found even before the copper oxides \cite{Ishiguro}.  The proximity of
SC to antiferromagnetism (AFM) in some, though not all (see below)
CTS, have led to theories of SC in the CTS that are related to their
counterparts in copper oxides
\cite{Vojta99a,Schmalian98a,Kino98a,Kondo98a,Powell05a,Gan05a,Sahebsara06a,Watanabe06a,Hebert15a}.
Furthermore, the occurrence of SC in the CTS at fixed carrier density
by application of pressure instead of doping has led to the belief
that understanding CTS may be easier and a first step towards
understanding correlated-electron SC itself. At the same time though,
occurrence of SC at fixed carrier concentration demands that there is
both broad and deep understanding of the role of this specific carrier
concentration, and this issue has therefore been at the center of the
discussions on organic SC, and is also the subject matter of the
present work.

\begin{figure}
  \centerline{\resizebox{3.0in}{!}{\includegraphics{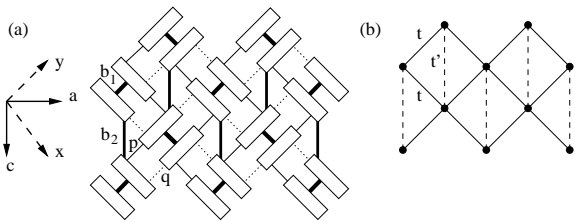}}}
  \caption{(a) Lattice structure of the ET layer in
    $\kappa$-(ET)$_2$X. The hopping integrals $t_{\rm b1}$, $t_{\rm b2}$,
    $t_{\rm p}$, and $t_{\rm q}$ are indicated. (b) The effective
    $\frac{1}{2}$-filled lattice: each dimer is now one lattice
    site. The effective hopping integrals $t =(|t_{\rm p}| + |t_{\rm q}|)/2$ and
    $t^\prime = t_{\rm b2}/2$.}
  \label{fig-lattice}
\end{figure}

Two fundamentally different approaches to correlated-electron
superconductivity in organic CTS exist in the literature. The first of
these is specific to strongly dimerized quasi-two dimensional
(quasi-2D) systems such as $\kappa$-(BEDT-TTF)$_2$X (hereafter
$\kappa$-(ET)$_2$X) and R[Pd(dmit)$_2$]$_2$ salts (here X and R are
monovalent counteranion and countercation, respectively), in which
pairs of cationic ET or anionic Pd(dmit)$_2$ molecules are strongly
coupled as dimers, and the dimers form anisotropic triangular lattices
(see Fig.~\ref{fig-lattice}) \cite{Kanoda11a}.  The charge on
individual molecules is 0.5 and hence the dimer lattice has been
considered as an {\it effective} $\frac{1}{2}$-filled band with strong
intradimer onsite Coulomb repulsion $U_{\rm eff}$.  Theoretical
discussions are within the anisotropic triangular lattice Hubbard
$U_{\rm eff}$-$t$-$t^\prime$ Hamiltonian, with hopping integrals $t$
along {\bf x} and {\bf y}, and $t^\prime$ along the diagonal {\bf y-x}
direction (see Fig.~\ref{fig-lattice}(b))
\cite{Kino98a,Schmalian98a,Kondo98a,Vojta99a,Baskaran03a,Liu05a,Kyung06a,Yokoyama06a,Watanabe06a,Sahebsara06a,Nevidomskyy08a,Sentef11a,Hebert15a}.
Metal-insulator transition occurs at a finite $U^c_{\rm eff}$, with
$U^c_{\rm eff}$ increasing with $t^\prime$. For relatively small
$|t^\prime/t|$ the semiconducting state is AFM, while for moderate to
large $t^\prime$ spin-liquid (SL) behavior is predicted. The theory
gives satisfactory understanding of the magnetic behavior of CTS with
small to moderate $|t^\prime/t| < 1$, including AFM in
$\kappa$-(ET)$_2$Cu[N(CN)$_2$]Cl (hereafter $\kappa$-Cl) with
$|t^\prime/t| \sim 0.44$ and apparent SL behavior in
$\kappa$-(ET)$_2$Cu$_2$(CN)$_3$ (hereafter $\kappa$-CN) and
$\beta^\prime$-EtMe$_3$Sb[Pd(dmit)$_2$]$_2$ with $|t^\prime/t| \sim
0.8-0.9$ \cite{Kanoda11a,Kandpal09a}. Very recent experimental work
\cite{Yamamoto17a} has concluded that the the simple SL picture as
well as the $\frac{1}{2}$-filled band Hubbard Hamiltonian are not
appropriate for $\beta^\prime$-EtMe$_3$Sb[Pd(dmit)$_2$]$_2$, a point
to which we return later.  Various mean field and dynamical mean field
theory calculations find $d$-wave SC within the $\frac{1}{2}$-filled
$U_{\rm eff}$-$t$-$t^\prime$ model at intermediate $|t^\prime/t|$,
bounded by an antiferromagnetic semiconductor and a correlated metal in the phase
diagram
\cite{Kino98a,Schmalian98a,Kondo98a,Vojta99a,Baskaran03a,Liu05a,Kyung06a,Yokoyama06a,Watanabe06a,Sahebsara06a,Nevidomskyy08a,Sentef11a,Hebert15a}.
Numerical investigations by us \cite{Clay08a,Dayal12a} and others
\cite{Watanabe08a,Shirakawa17a} have, however, demonstrated that
nonzero $U_{\rm eff}$ suppresses the long-range component of the
superconducting pair-pair correlations, indicating absence of SC
within the model. Additionally, the effective $\frac{1}{2}$-filled
band approach is inapplicable to CTS in which the organic monomer
molecules are not coupled as dimers, and charge order (CO) and not
AFM is proximate to SC
\cite{Kimura06a,Kobayashi86a,Tajima02a,Yamamoto11a,Shikama12a}.

An alternative theory \cite{Gomes16a,DeSilva16a,Clay19a} ascribes each
of the competing broken symmetries to the geometrically frustrated
{\it monomer} lattice of organic molecules with carrier density
$\rho=0.5$ per molecule and bandfilling $\frac{1}{4}$.  Although AFM
and Mott insulating behavior are expected in the strongly dimerized
and weakly frustrated $\rho=0.5$ band, for stronger lattice
frustration or weaker dimerization there is also a proximate static or
dynamically fluctuating CO state, driven by quantum effects rather
than classical electrostatics, that is a Wigner crystal of {\it
  spin-singlet pairs} (a paired electron crystal, hereafter PEC)
\cite{Li10a,Dayal11a}.  Uniquely to $\rho=0.5$, it is possible to have
a CO of spin-singlet pairs separated by pairs of vacant sites.
Dipolar-spin coupling \cite{Hotta10a} and polar charge fluctuations
\cite{Sekine13a} are related to dynamic CO fluctuations.  Within the
correlated $\frac{1}{4}$-filled band theory, pressure-induced
increased lattice frustration drives correlated motion of the
spin-pairs into the ``paired'' vacant sites, giving a wavefunction
that is still paired and constitutes the superconducting state
\cite{Gomes16a,DeSilva16a,Clay19a}. Our theory is at the interface of
the resonating valence bond (RVB) theory \cite{Anderson87a} and the
oldest version of the so-called bipolaron theory of superconductivity
\cite{Chakraverty85a}. The spin-singlet bonds here arise from antiferromagnetic
spin-coupling \cite{Gomes16a,DeSilva16a}, as opposed to overly strong
electron-phonon couplings, as assumed in the earlier literature
\cite{Chakraverty85a}.  We refer to this as the valence bond theory of
SC, distinct from the RVB theory by its focus on $\rho=0.5$.

The PEC concept is related to conjectures of ``density wave of Cooper
pairs'' in the high T$_c$ copper oxides
\cite{Franz04a,Tesanovic04a,Chen04a,Vojta08a,Hamidian16a,Cai16a,Mesaros16a},
where the precise nature of the density wave is still being debated
(see however, Reference \onlinecite{Mazumdar18a}).  AFM
is not the driver of SC within this second approach; SC in
$\kappa$-(ET)$_2$X is preceded by an ``orbital reordering'' that
renders unequal charge densities on the monomers within a dimer unit,
with mobile spin-singlet bonds forming between the charge-rich
molecules of neighboring dimers
\cite{Gomes16a,DeSilva16a,Clay19a,Sekine13a}. 
  Importantly, PEC formation is not limited to the dimerized
  $\kappa$-(ET)$_2$X, and the CO phases adjacent to SC in undimerized
  $\alpha$- $\beta$- and $\theta$-(ET)$_2$X are PECs \cite{Clay19a}.
  This, in turn, indicates that the mechanism of SC in the dimerized
  and undimerized CTS is the same.  Other calculations that have
found SC within the $\frac{1}{4}$-filled band Hubbard model using
mean-field and random phase approximations
\cite{Kondo01a,Kuroki06a,Sekine13a,Guterding16a}, or variational Monte
Carlo \cite{Watanabe19a} are not filling-specific; a key aspect of our
theory is that the PEC, and hence SC, are absent for bandfilling even
modestly far from $\frac{1}{4}$.

In the present paper we discuss a new class of $\kappa$-CTS that are
only now beginning to be investigated experimentally.  These materials
belong to the class $|t^\prime/t| > 1$ and allow sharp comparisons
between the two theoretical starting points described above, as we
point out in the present work. We show that while within the effective
$\frac{1}{2}$-filled band theory neither the observed strange
semiconductors nor superconductivity are expected, the ground state of
the $\frac{1}{4}$-filled Hubbard model is a strongly correlated metal
that is susceptible to all the observed exotic transitions.

\section{New Quasi-1D CTS and Theoretical Challenges}

Until recently, SC in the $\kappa$-(ET)$_2$X family was limited to
materials which lie firmly in the $|t^\prime/t| < 1$ region of the
effective $\frac{1}{2}$-filled Hubbard model. Discoveries of exotic
behavior including SC in $\kappa$ compounds with effective
$|t^\prime/t|$ significantly greater than 1 provide new testing
grounds for the two classes of theories.  The principal observations
are given below.

(i) SL behavior has perhaps been seen in $\kappa$-(ET)$_2$TaF$_6$
(hereafter $\kappa$-TaF$_6$) \cite{Kawamoto18a}, with $|t^\prime/t| >
1.5$. Absence of magnetic ordering has been demonstrated down to 1.6 K
\cite{Kawamoto18a}.

(ii) $\kappa$-(ET)$_2$B(CN)$_4$ (hereafter $\kappa$-B(CN)$_4$) is an
insulator at ambient pressure with effective $|t^\prime/t| \sim 1.8$
\cite{Yoshida15a}.  It undergoes phase transition into a nonmagnetic
state with spin-gap (SG) \cite{Yoshida15a} at $\sim 5$ K. Raman
measurements have failed to find CO so far, but the
experiments were done only down to to 10 K.

(iii) $\kappa$-(ET)$_2$CF$_3$SO$_3$ (hereafter $\kappa$-CF$_3$SO$_3$)
contains two inequivalent ET layers, labeled layers A and B, separated
by the anion layer \cite{Fettouhi95a,Ito16a}. Within the effective
model $|t^\prime/t|=1.48$ (1.78) for layer A (B) at low temperatures
\cite{Ito16a}.  Following structural changes at 230 K and 190 K, AFM
 occurs at ambient pressure at a low N\'eel temperature T$_{\rm
  N}$ of 2.5 K \cite{Ito16a}.  Transition to a metallic state and SC
occurs under pressure (T$_{\rm c,max}$ = 4.8 K at 1.3 GPa)
\cite{Ito16a}.  T$_{\rm N} <$ T$_{\rm c}$ (albeit pressure-induced) is
unknown in any other correlated electron superconductor that has shown
transition from AFM to SC.

The above observations defy explanations within the effective
$\frac{1}{2}$-filled band approach. First, even as magnetic
measurements on $\kappa$-TaF$_6$ are awaited at lower temperatures ($
< 1.6$ K), SL behavior is not expected within the effective
$\frac{1}{2}$-filled band model for $|t^\prime/t| \geq 1.5$. Theory
limits the SL phase only to $|t^\prime/t| \leq 1$ within the
triangular lattice $\frac{1}{2}$-filled Hubbard Hamiltonian
\cite{Yoshioka09a,Tocchio14a,Acheche16a}.  What is perhaps more
relevant is that very recent experimental works
\cite{Itoh13a,Yakushi15a,Yamamoto17a,Fujiyama18a} have cast doubt on
whether even $\kappa$-CN and EtMe$_3$Sb[Pd(dmit)$_2$]$_2$, which were
thought to be prime candidates for SL in the parameter region
$|t^\prime/t| \leq 1$ until very recently \cite{Kanoda11a}, should
really be described as such. The original proponents of the SL picture
for EtMe$_3$Sb[Pd(dmit)$_2$]$_2$, in particular, have found that the
driver of the geometrical frustration here are hidden {\it charge and
  lattice fluctuations}, not expected within the simple
$\frac{1}{2}$-filled band Hubbard model \cite{Yamamoto17a}.
  
Second, within the effective $\frac{1}{2}$-filled band model, the SG
transition observed in $\kappa$-B(CN)$_4$ can only be due to
spin-Peierls transition, or due to a frustration-driven transition to
a valence bond solid (VBS).  The large ratio of inter-chain to
intra-chain hopping, $\sim$ 0.6 in $\kappa$-B(CN)$_4$, implies absence
of 1D nesting necessary for the spin-Peierls transition characteristic
of quasi-1D systems (note here that spin-density wave, and not the
spin-Peierls state characterizes the spatial broken symmetry in
(TMTSF)$_2$X where the same ratio at 0.2 is significantly smaller).
It is also known from previous theoretical work that the Hubbard
repulsion severely reduces bond dimerization in the 2D half-filled
band even where nesting would have permitted this in the uncorrelated
limit \cite{Mazumdar87a,Pillay13a}.  The VBS is also precluded
theoretically within the triangular lattice $\frac{1}{2}$-filled band
Hubbard model \cite{Gomes13a}.  We therefore conclude that that the
explanation of the SG transition in $\kappa$-B(CN)$_4$ requires going
beyond the effective $\frac{1}{2}$-filled band model.

Finally, the observed SC in $\kappa$-CF$_3$SO$_3$ also lies outside
the domain of applicability of the effective $\frac{1}{2}$-filled band
theories.  SC appears for significantly smaller $|t^\prime/t| < 1$
within these theories
\cite{Kino98a,Schmalian98a,Kondo98a,Vojta99a,Baskaran03a,Liu05a,Kyung06a,Yokoyama06a,Watanabe06a,Sahebsara06a,Nevidomskyy08a,Sentef11a,Hebert15a}.
In the present paper we show from explicit numerical calculations
based on the competing $\frac{1}{4}$-filled band Hubbard model that
the ground state of the latter is a strongly-correlated metal in which
none of the above exotic phases are precluded, and hence any one of
these can presumably dominate when small but nonzero interactions
excluded in the purely electronic Hamiltonian are
included. Superconducting correlations, in particular, are enhanced by
the Hubbard $U$ in the $\frac{1}{4}$-filled band, relative to the
noninteracting $U=0$ limit, which is a necessary condition for
correlated-electron SC.

\section{Theoretical model, parameters and computational techniques}
\label{model}

We consider the Hubbard Hamiltonian for the $\kappa$ lattice
structure of Fig.~\ref{fig-lattice}(a),
\begin{equation}
  H = \sum_{\langle ij\rangle,\sigma}t_{ij}(c^\dagger_{i,\sigma}c_{j,\sigma}+H.c.)
  + U\sum_i n_{i,\uparrow}n_{i,\downarrow},
  \label{ham}
\end{equation}
where $c^\dagger_{i,\sigma}$ ($c_{i,\sigma}$) creates (annihilates) an
electron of spin $\sigma$ on the highest molecular orbital of a {\it
  monomer} ET molecule $i$ (hereafter ``site''), and all other terms
have their usual meanings.  In our calculations for
$\kappa$-CF$_3$SO$_3$ we take hopping integrals ($t_{\rm b1}$, $t_{\rm
  b2}$, $t_{\rm p}$, $t_{\rm q}$; $|t^\prime/t|$) in eVs to be (0.234, 0.157, 0.029,
-0.077; 1.48) and (0.248, 0.169, 0.048, -0.047; 1.78), for layers A and B,
respectively \cite{Ito16a}. We report separate calculations for
$\kappa$-B(CN)$_4$ with corresponding hopping parameters
\cite{Yoshida15a} (0.221, 0.117, 0.035, -0.029; 1.83).  These parameters
make $\kappa$-B(CN)$_4$ closer to 1D than $\kappa$-CF$_3$SO$_3$, as
already noted above.  These hopping integrals are for electrons and
not holes; the corresponding carrier density is $\rho=1.5$.  Our zero
temperature computational results are for finite periodic clusters
with 32, 64, and 128 monomer sites \cite{Supplemental} using the Path Integral
Renormalization Group \cite{Kashima01b,Mizusaki04a} and Constrained
Path Monte Carlo \cite{Zhang97a} methods \cite{Supplemental}.
For both methods we fully incorporate lattice and spin-parity
  symmetries (see References \onlinecite{Mizusaki04a} and \onlinecite{Shi14a}), which
  significantly increases their accuracy. Our prior benchmark results
  of these methods  are available in
  References \onlinecite{Dayal12a}, \onlinecite{Gomes16a}, and \onlinecite{DeSilva16a}.

\section{Computational results}

\subsection{Magnetic behavior}
\label{magnetic} 
\subsubsection{Weak AFM in $\kappa$-CF$_3$SO$_3$}
\begin{figure}
  \begin{center}
    \begin{overpic}[width=1.5in]{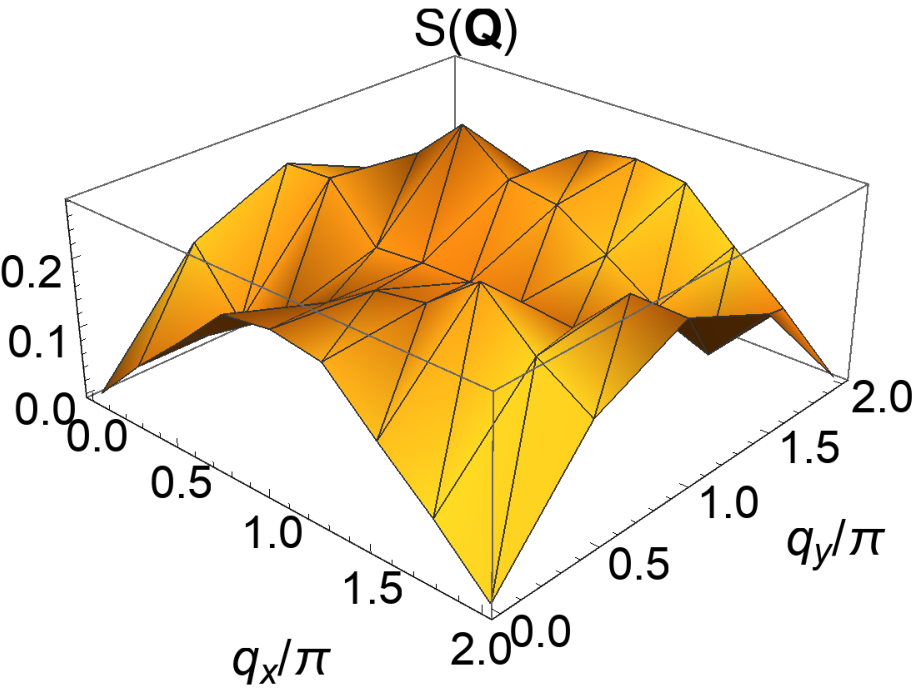}
      \put (0,80) {\small(a)}\hspace{0.1in}
    \end{overpic}\hspace{0.1in}%
    \begin{overpic}[width=1.5in]{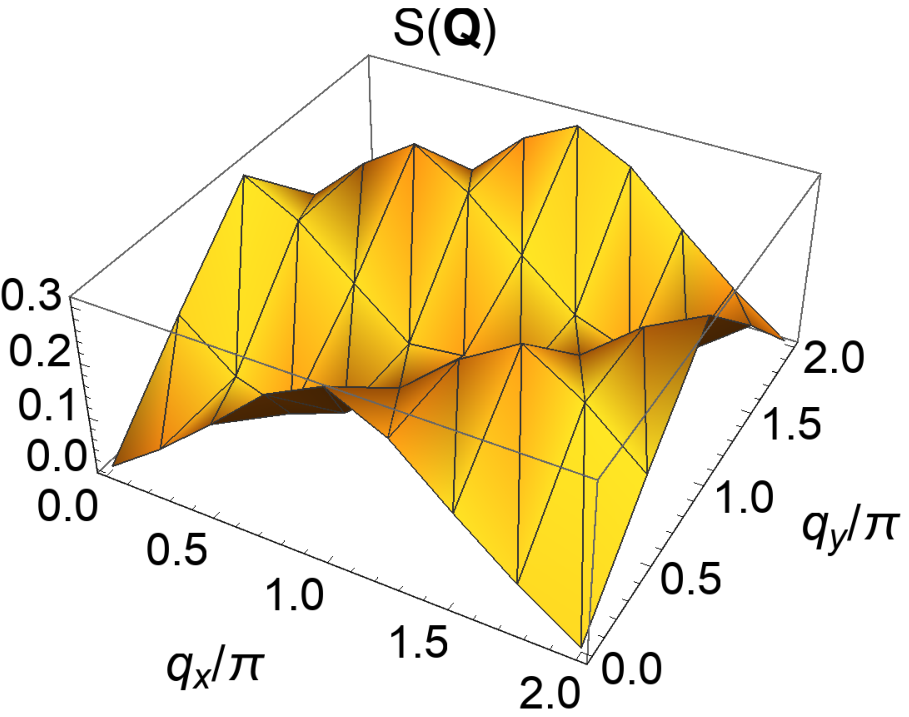}
      \put (0,80) {\small(b)}\hspace{0.1in}
    \end{overpic}
    \begin{overpic}[width=1.5in]{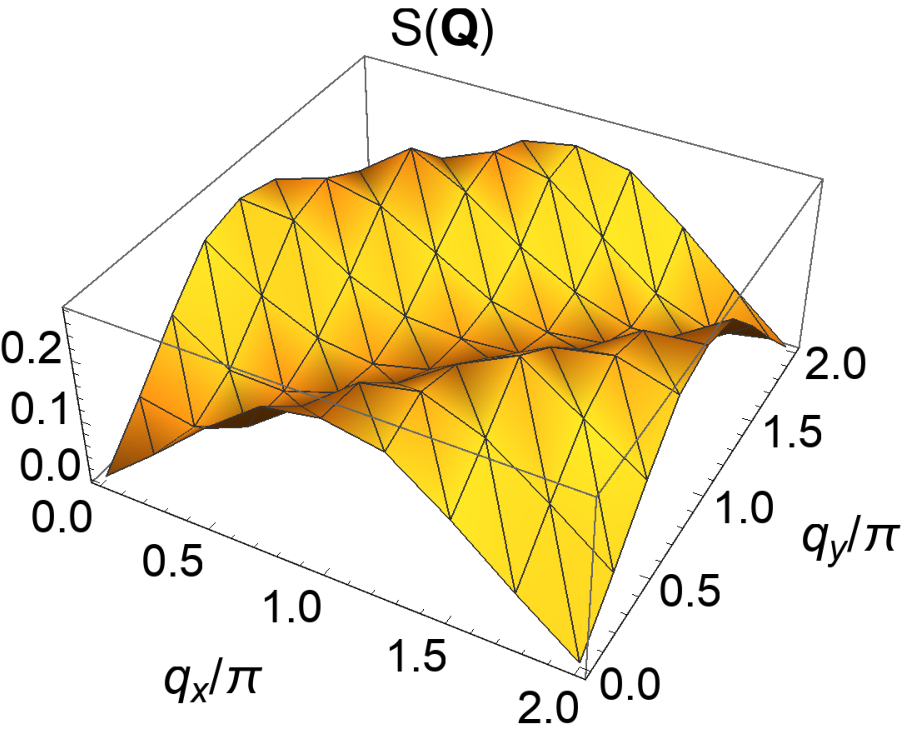}
      \put (0,80) {\small(c)}\hspace{0.1in}
    \end{overpic}\hspace{0.1in}%
    \begin{overpic}[width=1.5in]{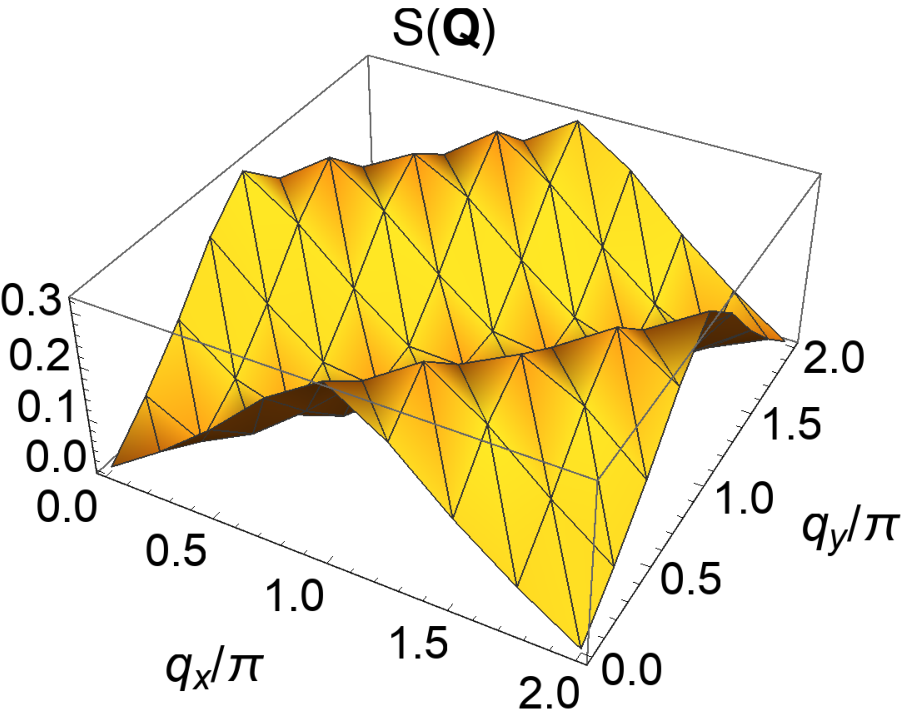}
      \put (0,80) {\small(d)}\hspace{0.1in}
    \end{overpic}
  \end{center}
  \caption{(color online) Dimer spin structure factor S({\bf Q}) for
    the $\kappa$-CF$_3$SO$_3$ lattice with
    $\rho=1.5$ and $U$=0.5 eV.  (a) 64 sites, layer A, (b) 64 sites,
    layer B, (c) 128 sites, layer A, and (d) 128 sites, layer B.}
  \label{AFM-CF3SO3}
\end{figure}  
\begin{figure}
  \begin{center}
    \resizebox{3.3in}{!}{\includegraphics{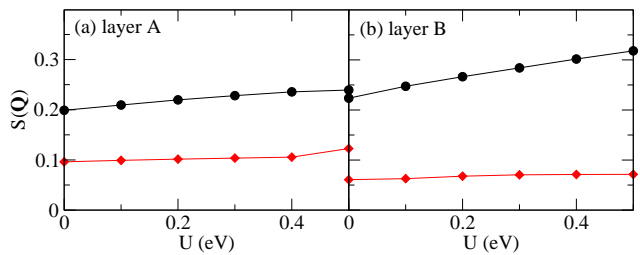}}
  \end{center}
  \caption{(color online)
$S({\bf Q})$ versus $U$ for a 128-site $\kappa$-CF$_3$SO$_3$ cluster.
    Circles and diamonds are for {\bf Q}=($\pi$,0) and ($\pi$,$\pi$)
    respectively.}
  \label{peaks}
\end{figure}

The tendency to AFM is best discussed within the dimer
representation \cite{DeSilva16a}.  We define the total $z$-component
of the spin on dimer $i$ as
\begin{equation}
  S_i^z  = \frac{1}{2} (n_{i_1,\uparrow} + n_{i_2,\uparrow}
  - n_{i_1,\downarrow} - n_{i_2,\downarrow}).
  \label{dimerspin}
\end{equation}
In Eq.~\ref{dimerspin}, $i_1$ and $i_2$ refer to the two different
molecules within the dimer $i$ and
$n_{j,\sigma}=c^\dagger_{j,\sigma}c_{j,\sigma}$. We calculate the
dimer spin structure factor,
\begin{equation}
  S({\bf Q})=\frac{1}{N_d} \sum_{j,k} e^{i{\bf Q}\cdot({\bf r_j}-{\bf r_k})}
  \langle S^z_j S^z_k\rangle,
  \label{sfac}
\end{equation}
where $N_d$ is the number of dimers and dimer position vectors {\bf
  r$_j$} are defined on a conventional square lattice, whose {\bf x}
and {\bf y} axes are indicated in Fig.~\ref{fig-lattice}.
Fig.~\ref{AFM-CF3SO3} shows the {\bf Q} dependence at $\rho=1.5$ of
S({\bf Q}) for 64 and 128 monomer lattice sites, separately for layers
A and B in each case.  The {\bf Q} dependence of $S({\bf Q})$ here is
similar to that of the $\frac{1}{2}$-filled Hubbard
model\cite{Tocchio14a} with $|t^\prime/t|>1$.  We find a line of
maxima in $S({\bf Q})$ perpendicular to the $t^\prime$ direction, with
the dominant wavevector $\pi$ along each chain.  $S({\bf Q})$ does not
increase with system size, indicating the absence of long-range order.
Fig.~\ref{peaks} shows that $S({\bf Q})$ at ($\pi,0$) is stronger than
that at ($\pi,\pi$) in agreement with results for the
$\frac{1}{2}$-filled Hubbard model
\cite{Clay08a,Tocchio14a,Acheche16a}.  $S(\pi,\pi)$ here is nearly
half of that calculated by us previously for $\kappa$-Cl
\cite{DeSilva16a}. The dominance of $S(\pi,0)$ over $S(\pi,\pi)$ is a
consequence of quasi-1D character. Note however that in the 1D limit
there is no long-range AFM.  Additionally, the absence of true
long range order in $S({\bf Q})$ in our calculations indicates that
$\kappa$-CF$_3$SO$_3$ is barely insulating and is close to a
correlated metallic phase.  The enhanced 1D character and the
proximity to the metallic state, taken together, are the likely
reasons for the small T$_{\rm N}$ in $\kappa$-CF$_3$SO$_3$.
\begin{figure}
  \begin{center}
    \raisebox{0.1in}{
      \resizebox{1.0in}{!}{\includegraphics{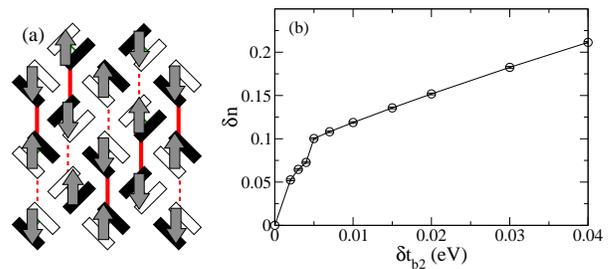}}}\hspace{0.1in}%
    \resizebox{2.0in}{!}{\includegraphics{fig4b}}
  \end{center}
  \caption{(a) (color online) PEC accompanying bond alternation of
    $t_{b2}$ bonds. Introducing the strong (solid red) and weak (dashed
    red) $t_{b2}$ bonds leads to the indicated CO, where black and white
    boxes represent charge-rich and charge-poor ET molecules,
    respectively \cite{Supplemental}. Spins superimposed on the dimers represent the
    magnetic ordering in the ($\pi$,0) antiferromagnetic phase.  (b) 128 site CO
    order parameter, the difference in charge density between the
    charge-rich and the charge-poor molecules as a function of the
    difference in the $t_{\rm b2}$ hopping integrals. Calculations are
    for hopping parameters for $\kappa$-B(CN)$_4$ with $U$=0.4 eV.}
 \label{PEC}
\end{figure}

\subsubsection{Transition to the spin-gapped state in $\kappa$-B(CN)$_4$}

As mentioned above, simple spin-Peierls transition or VBS formation,
as have been suggested within the effective $\frac{1}{2}$-filled model
\cite{Yoshida15a}, cannot be the origin of the observed SG in
$\kappa$-B(CN)$_4$ \cite{Mazumdar87a,Pillay13a,Gomes13a}.  The recent
demonstration that the driving forces behind the apparent SL-like
behavior in $\beta^\prime$-EtMe$_3$Sb[Pd(dmit)$_2$]$_2$ are charge and
lattice fluctuations \cite{Yamamoto17a}, which are precluded in the
$\frac{1}{2}$-filled band but are {\it expected} within the
$\frac{1}{4}$-filled band characterization of the system
\cite{Clay19a}, gives a hint to the mechanism of the SG in
$\kappa$-B(CN)$_4$.  Since similar intradimer charge inequalities can
occur in all $\frac{1}{4}$-filled band systems, we investigate the
likelihood of PEC formation in the present case.

We have performed numerical calculations of charge densities in which
the $t_{\rm b2}$ hopping integrals (see Fig.~\ref{fig-lattice})
alternate as $t_{\rm b2}^0 \pm \delta t_{\rm b2}$, as suggested in
Reference \onlinecite{Yoshida15a}. All other hopping parameters 
correspond to those for $\kappa$-B(CN)$_4$, given in Section
\ref{model} above. The PEC charge order pattern (Fig.~\ref{PEC}(a))
appears {\it spontaneously} for nonzero $\delta t_{\rm b2}$, as shown
in Fig.~\ref{PEC}(b), with CO $\cdots0110\cdots$ along the bonds in
the {\bf y} and {\bf x}-{\bf y} directions, and $\cdots1010\cdots$
along the bonds in the third direction {\bf x} (see
Fig.~\ref{fig-lattice}(a) and Reference \onlinecite{Supplemental}).
Here `1' and `0' represent charge-rich and charge-poor molecules,
respectively.  Period doubling along both the crystal $a$ and the
$c$-axes are therefore expected at the SG transition, as opposed to
only along $t_{\rm b2}$ \cite{Yoshida15a}.  Note that the pattern and
phase of the CO is consistent with ($\pi,0$) periodicity of the AFM,
if we make the reasonable assumption of strong antiferromagnetic correlations
between nearest neighbor charge-rich molecules.

Fig.~\ref{PEC}(b) gives the calculated CO order parameter $\delta n$,
the difference in the charge densities between the charge-rich and the
charge-poor molecules.  The kink at $\delta t_{\rm b2}\sim0.005$ eV is
due to a level crossing; for $\delta t_{\rm b2}>0.005$ eV the
difference in $\delta$n for 64 versus 128 sites is less than the size
of the symbols in Fig.~\ref{PEC}(b).  We consider the range of $\delta
t_{\rm b2}$ in Fig.~\ref{PEC}(b) to be realistic \cite{Yoshida15a},
and while the calculated $\delta n$ are small, they should be
observable if experiments sensitive to the molecular charge are
extended to T$<$T$_{\rm SG}$.  Even more importantly, we predict
period doubling along {\it both} {\bf a} and {\bf c}-directions,
consistent with PEC formation, that should also be experimentally
verifiable.

\subsection{Superconductivity: $\kappa$-CF$_3$SO$_3$ versus $\kappa$-B(CN)$_4$}
\label{sc}

\subsubsection{Coulomb enhancement of superconducting correlations: $\kappa$-CF$_3$SO$_3$}

We now probe the question of SC. As in previous work
\cite{DeSilva16a}, calculations here are for variable carrier
densities $\rho$, since the basic contention of our theory
\cite{Gomes16a,DeSilva16a,Clay19a} is that SC is specific to $\rho$
exactly equal to or close to 1.5, and is absent in $\rho$ even
modestly far from 1.5.  We calculate superconducting pair-pair
correlations $P_{ij}=\langle \Delta_i^\dagger \Delta_j \rangle$, where
$\Delta_i^\dagger$ creates a superposition of singlet pairs between
monomer sites belonging to neighboring dimers. Such interdimer singlet
pairs necessarily create unequal intradimer charge densities
\cite{DeSilva16a}.
\begin{figure}
  \begin{center}
    \resizebox{3.2in}{!}{\includegraphics{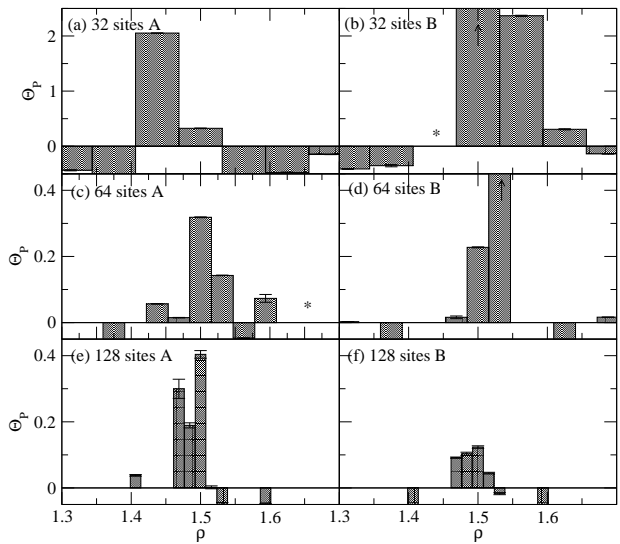}}
  \end{center}
  \caption{The enhancement factor $\Theta_P$ for the long-range
    component of the pair-pair correlations with $d_2$ ($s$ + $d_{x^2-y^2}$)
    symmetry, for hopping integrals corresponding to the
    $\kappa$-CF$_3$SO$_3$ lattice. $\Theta_P > 0$ implies pair-pair
    correlations enhanced over their $U=0$ values. Pair-pair
    correlations are enhanced for a narrow region of carrier density
    $\rho=1.5$ and are suppressed elsewhere (see text).  Densities
    marked with * were removed due to finite-size effects
    \cite{Supplemental}.  The computations for the 64- and 128-site
    clusters were done for density ranges narrower (1.3 - 1.7 and 1.4
    - 1.6, respectively) than for the 32-site cluster based on
    previous work for $|t^\prime/t|<1$ that had shown peaking of
    $\Theta_P$ at $\rho=1.5$.}
\label{pairing}
\end{figure}
Considering singlet pairs between monomer molecules on a central dimer
and its six neighboring dimers, there occur four types of $d$-wave
pairings (see Fig.~3 in Reference \onlinecite{DeSilva16a}).  The first
of these pairing symmetries, $d_1$, is similar to the usual
$d_{x^2-y2}$ symmetry and the second, $d_2$, is a form of
$s+d_{x^2-y^2}$ pairing also considered by other authors
\cite{Guterding16a,Guterding16b,Watanabe17a}.  For all three clusters
we calculate the average long-range pair-pair correlations $\bar{P} =
N_P^{-1} \sum_{|r_{ij}|>2}P_{ij}$, where $N_P$ is the number of terms
in the sum with the restriction $r_{ij}>2$ in units of dimer-dimer
spacing \cite{DeSilva16a}.  Since the minimal requirement for
interaction-driven SC is that nonzero Hubbard $U$ {\it enhances} the
pairing correlations, we calculate the enhancement factor $\Theta_P
=[\bar{P}(U)/\bar{P}(U=0)]-1$. Then $\Theta_P > 0$ implies likelihood
of pairing and $\Theta_P < 0$ indicates suppression of pairing by Hubbard
$U$.  Previous calculations \cite{DeSilva16a} for effective
$|t^\prime/t| < 1$ found enhancement only for $\rho \sim 1.5$.  A
precise value for $U$ for the actual materials is difficult to
estimate.  In the quasi-1D (TMTSF)$_2$X materials, $U \sim $ 1 eV has
been estimated for the extended Hubbard model \cite{Mila95a}. Based on
the larger size of the ET molecule we expect a smaller $U$ in the
range 0.4 -- 0.8 eV in $\kappa$-(ET)$_2$X compared to (TMTSF)$_2$X.

\begin{figure}
  \begin{center}
    \resizebox{3.2in}{!}{\includegraphics{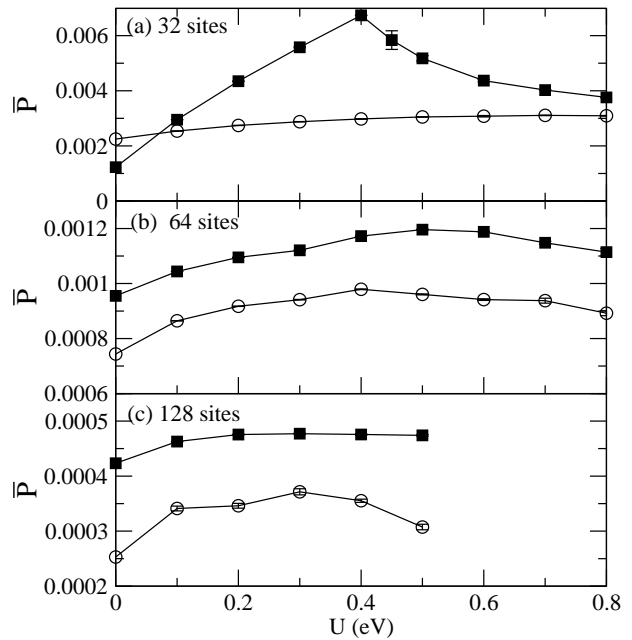}}
  \end{center}
  \caption{Average long-range pair-pair correlation $\bar{P}$ with
    $d_2$ ($s$ + $d_{x^2-y^2}$) symmetry versus $U$ for hopping
    parameters corresponding to $\kappa$-CF$_3$SO$_3$ at $\rho=1.5$.
    Open (filled) symbols are for layer A (B).}
\label{U-dependence}
\end{figure}
Our calculations find very strong enhancements of pair correlations
with $d_2$ symmetry for $\rho \sim 1.5$ for both layers A and B of
$\kappa$-CF$_3$SO$_3$, for all three cluster sizes ($d_2$ gave the
most strongly enhanced pair correlations \cite{DeSilva16a} also for
$\kappa$-Cl and $\kappa$-CN).  In Fig.~\ref{pairing} we show plots of
$\Theta_P(d_2)$ versus $\rho$ for all three cluster sizes, for $U=0.4$
eV, separately for layers A and B in each case (see Supplementary
Materials \cite{Supplemental} for results on $\Theta_P(d_1)$).  The
calculated results are for a wide range of $\rho$ in the 32 and
64-site clusters, spanning $\rho=1.3-1.7$. A narrower density range
$1.4-1.6$ was considered for the 128-site cluster, based on the
similar behavior of the three clusters.  In spite of the quasi-1D
character of the lattice, here we find nearly identical results as in
the earlier work for the 2D lattice, where an even broader density
range $\rho=1.0-2.0$ was considered \cite{DeSilva16a}.  In all cases
we find that $\Theta_P(d_2) > 0$ occurs over a very narrow density
range about 1.5.  In Figs.~\ref{pairing}(c) and (e) we see dips
followed by rise in $\Theta_P$ for densities close to but slightly
away from density exactly 1.5. However, these dips and increases are
tiny compared to the large peaks at density exactly at or very close
to 1.5.  The widths of the vertical shaded bars in Fig.~\ref{pairing}
decrease with increasing lattice size, since between any two densities
there are fewer total electron numbers in the smaller lattices.
Importantly, (i) $\Theta_P(d_2) < 0$  outside of the density range
1.5 $\pm$ 0.1 in lattices larger than 32 sites, (ii) for all cases the
largest enhancement in the pairing correlations is found either for
$\rho$ exactly equal to 1.5 or for an electron number immediately larger or
smaller for a given lattice, and (iii) the range of densities over
which pair correlations are enhanced by $U$ {\it decreases} with
increasing cluster size.  Quantitatively speaking, the enhancements of
the pair correlations for $\rho=1.5$ here are either larger than (for
32 sites) or comparable to (for 64 sites) \cite{DeSilva16a} those for
effective $|t^\prime/t| < 1$.  Calculations for the very large
128-site clusters had not been performed before.

For the specific case of $\rho = 1.5$ we also performed calculations
of the enhancement factor over a broad range of $U$ for the 32- and
64-site clusters (see Fig.~\ref{U-dependence}).  For the 32 site layer
B lattice a quantum phase transition occurs at $U\approx$ 0.4 eV (see
Fig.~\ref{U-dependence}(a)).  In this lattice at $U\approx0.4$ eV a
sharp increase in $S(\pi,\pi)$ coupled with a decrease in $\bar{P}$
indicates a transition to an insulating antiferromagnetic state.  This is a
finite-size effect that is not present in larger lattices. On the
other hand, this result does indicate that our strongly-correlated
metal is close to an insulating phase, which would be likely a weak
AFM, based on our calculations of the magnetic structure factor, see
Section \ref{magnetic}.

\subsubsection{Reduced tendency to superconductivity in $\kappa$-B(CN)$_4$}

\begin{figure}
  \begin{center}
    \resizebox{3.2in}{!}{\includegraphics{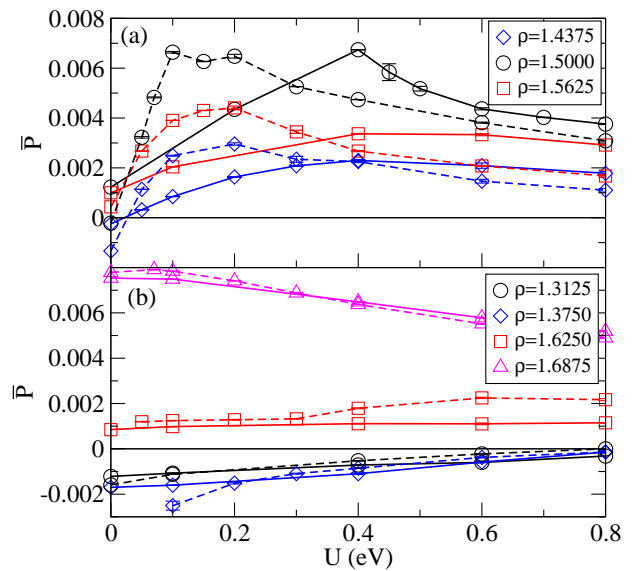}}
  \end{center}
  \caption{(color online) Average long-range pair-pair correlation
    $\bar{P}$ with $d_2$ ($s$ + $d_{x^2-y^2}$) symmetry versus $U$ for
    32-site lattices. Solid (dashed) lines correspond to parameters
    for $\kappa$-CF$_3$SO$_3$ ($\kappa$-B(CN)$_4$). Panels (a) and (b)
    show densities near to and further from $\rho=1.5$, respectively.}
\label{bcn4}
\end{figure}

 We have calculated $U$-dependent pair correlations also for the
 hopping parameters corresponding to $\kappa$-B(CN)$_4$, for clusters
 consisting of 32 monomer sites. We show these results in
 Figs.~\ref{bcn4}(a) and (b), where we have included the computational
 results for the 32-site $\kappa$-CF$_3$SO$_3$ lattice for
 comparison. The calculated pair-pair correlations are nearly
 identical for the two lattices for $\rho$ far from 1.5
 (Fig.~\ref{bcn4}(b)), where there is absence of enhancement, or even
 suppression of these correlations by the Hubbard $U$.  The results
 for $\rho \sim 1.5$ (Fig.~\ref{bcn4}(a)) are noticeably different,
 however.  On the one hand, the pair-pair correlations are enhanced by
 small Hubbard $U$ for $\kappa$-B(CN)$_4$. On the other, the
 enhancement occurs over a significantly smaller range of $U$ compared
 to that for $\kappa$-CF$_3$SO$_3$, with the peak in the enhancement
 occurring at very small ($\alt$0.2 eV) $U$. For realistic molecular
 $U$ ($>$ 0.4 eV) the predicted tendency to SC in $\kappa$-B(CN)$_4$
 is thus considerably smaller than in $\kappa$-CF$_3$SO$_3$.

\section{Conclusions}

In summary, we have performed large scale accurate
  numerical calculations within the Hubbard Hamiltonian, for quasi-1D
  triangular lattice clusters of three different sizes, corresponding
  to the monomer $\kappa$-(ET)$_2$X lattice of
  Fig.~\ref{fig-lattice}(a), with realistic parameter values for
  $\kappa$-CF$_3$SO$_3$ and $\kappa$-B(CN)$_4$.  Our results can be
  summarized as follows.

(i) For hopping parameters corresponding to both the inequivalent
  layers A and B of $\kappa$-CF$_3$SO$_3$, we find that S({\bf Q}) peaks
  at $(\pi,0)$ instead of $(\pi,\pi)$. This is in agreement with
  results obtained for the effective $\frac{1}{2}$-filled band Hubbard
  model for $|t^\prime/t| > 1$.

(ii) For hopping parameters corresponding to $\kappa$-B(CN)$_4$ we
  find the smallest alternation in $t_{b2}$, which has been suggested
  as the origin of the SG transition \cite{Yoshida15a}, is accompanied
  by CO as well as bond distortions in the two other directions,
  leading to period doublings.  This is the classic signature of a
  tendency to unconditional transition to the PEC, noted previously
  for the effective $t^\prime < t$ materials \cite{Li10a,Dayal11a}.

(iii) Returning to the hopping parameters corresponding to
  $\kappa$-CF$_3$SO$_3$, we find from electron density-dependent
  calculations on three different lattice sizes that superconducting
  pair-pair correlation functions of $d_2$ symmetry are enhanced by
  the Hubbard $U$ only for density close to 1.5, precisely the carrier
  density in superconducting CTS. The superconducting symmetry is thus
  the same \cite{DeSilva16a} as that for effective $|t^\prime/t| <
  1$. Also in agreement with previous calculations \cite{DeSilva16a},
  for electron densities even weakly away from 1.5 the Hubbard $U$
  suppresses pair correlations, indicating absence of SC here.

 (iv) For hopping parameters corresponding to $\kappa$-B(CN)$_4$,
  similar electron density-dependent calculations for the 32-site
  cluster find enhancement of superconducting pair-pair correlation
  functions for $\rho=1.5$, but only for $U$ values smaller than the
  known Hubbard $U$ for the ET molecule.  This explains the absence of
  SC to date in this system, but also suggests that application of
  pressure, which would increase the bandwidth and decrease the
  effective $U/|t|$, can induce SC here. This predicted result is
  reminiscent of the experimental observation of high pressure-induced
  superconducting transition \cite{Adachi00a} from the spin-Peierls
  state in (TMTTF)$_2$PF$_6$, which is a PEC at the quasi-1D limit
  \cite{Clay19a}.

The overall conclusion is that except for the weaker tendency to
metal-insulator transition and AFM within the purely
electronic Hamiltonian, which presumably is the reason behind the
small T$_{\rm{N}}$ in $\kappa$-CF$_3$SO$_3$, all other behaviors at
low temperatures are the same as what would be expected for the
effective $|t^\prime/t| < 1$.  The fundamental reason for this is that
the intradimer charge degrees of freedom become important at low
temperatures in both $\kappa$-(ET)$_2$X and R[Pd(dmit)$_2$]$_2$. In
recent years investigators have noted from different experimental
studies on effective $|t^\prime/t| < 1$ dimerized CTS that the lowest
excitations in these involve intradimer charge fluctuations and the
resultant lattice fluctuations \cite{Itoh13a,Yakushi15a,Fujiyama18a}.
Here we note that the same is true for effective $|t^\prime/t| >
1$. Should experiments at T $< 1.6$ K confirm the absence of magnetic
long-range order in $\kappa$-TaF$_6$, this should be ascribed also to
intradimer charge fluctuations. The SG transition in
$\kappa$-B(CN)$_4$ is likely due to PEC formation, as in the
$|t^\prime/t| < 1$ compound $\beta$-EtMe$_3$P[Pd(dmit)$_2$]$_2$
\cite{Yamamoto17a,Clay19a}. We suggest experiments sensitive to CO in
$\kappa$-B(CN)$_4$.

Finally, the observation of SC in $\kappa$-CF$_3$SO$_3$ is a strong
proof of the validity of our valence bond theory of SC.  Elsewhere
\cite{Clay19a} we have drawn attention to the very large number of
disparate families of inorganic correlated-electron superconductors
that share two features with CTS superconductors, viz., (a)
$\frac{1}{4}$-filling on frustrated lattices, and (b) SC proximate to
unconventional charge-density waves. The very large number of such
families suggests the universality of the proposed PEC-to-SC
mechanism. Indeed, one of us has suggested the same mechanism of SC
also for high T$_c$ copper oxides, both hole and electron-doped, where
an effective correlated $\frac{1}{4}$-filled oxygen band is reached
following a dopant-induced Cu$^{2+} \to$ Cu$^{1+}$ valence transition
\cite{Mazumdar18a}.

\section{Acknowledgments}

S.M. acknowledges partial support from NSF-CHE-1764152.  Some
calculations in this work used the Extreme Science and Engineering
Discovery Environment \cite{xsede} (XSEDE), which is supported by
National Science Foundation grant number ACI-1548562. Specifically, we
used the Bridges system \cite{bridges} which is supported by NSF
award number ACI-1445606, at the Pittsburgh Supercomputing Center
(PSC) under award TG-DMR190052.

\end{document}